# Artificial Intelligence for 5G Wireless Systems: Opportunities, Challenges, and Future Research Directions

Youness Arjoune
School of Electrical Engineering and Computer Science
University of North Dakota
Grand Forks, ND-58202, USA
youness.arjoune@und.edu

Saleh Faruque
School of Electrical Engineering and Computer Science
University of North Dakota
Grand Forks, ND-58202, USA
saleh.faruque@und.edu

*Abstract*—The advent of the wireless communications systems augurs new cutting-edge technologies, including self-driving vehicles, unmanned aerial systems, autonomous robots, the Internet of-Things, and virtual reality. These technologies require high data rates, ultra-low latency, and high reliability, all of which are promised by the fifth generation of wireless communication systems (5G). Many research groups state that 5G cannot meet its demands without artificial intelligence (AI) integration as 5G wireless networks are expected to generate unprecedented traffic giving wireless research designers access to big data that can help in predicting the demands and adjust cell designs to meet the users' requirements. Subsequently, many researchers applied AI in many aspects of 5G wireless communication design including radio resource allocation, network management, and cyber-security. In this paper, we provide an in-depth review of AI for 5G wireless communication systems. In this respect, the aim of this paper is to survey AI in 5G wireless communication systems by discussing many case studies and the associated challenges, and shedding new light on future research directions for leveraging AI in 5G wireless communications.

## I. Introduction

5G Wireless communication and mobile networks are facing many challenges to meet the unprecedented growing demands for access to wireless services with ultra-low latency and high data rates. 5G network today is the core technology of many cutting-edge technologies such as the internet-of-things (IoT), smart grid, unmanned aerial systems, and self-driving vehicles. 5G wireless networks are required to be characterized by high flexibility in design and resource management and allocation to meet the increasing demands of these heterogeneous networks and users.

The 5G specifications, released in 2017 by 3GPP, consider the flexibility of design as one but a fundamental pillar of 5G New Radio, which can be achieved through the integration of the software-defined network (SDN) and virtual network function (NVF) functions. Such flexibility permits a flexible 5G system that can adjust itself in real-time to optimize resource allocation while enhancing the quality of experience of users, which requires accurate prediction of the network behaviors, the traffic demands, and user's mobility. Many Wireless research leading groups predict that Artificial Intelligence (AI) is the next big "game-changing" technology, poised to provide 5G with the flexibility and the intelligence needed. for this reason, many researchers have investigated the efficiency of this theory in many aspects of 5G wireless communications including modulation, channel coding, interference management, and scheduling, 5G slicing, caching, energy efficiency, and cyber-security.

Many survey and tutorial papers provided an overview of wireless artificial intelligence. For instance, the authors of [1] provided a comprehensive tutorial on how deep learning can empower several applications in wireless systems. Specifically, the authors focused on some types of neural networks, such as recurrent, spiking, and deep neural networks, and how it can empower some wireless communication problems. Nevertheless, most of these survey papers focus on artificial intelligence theory, which is well-established, more than they focus on how this technology can solve practical problems in wireless communications. In this paper, we review wireless artificial intelligence with a particular focus on how AI methodologies can solve complex problems in 5G wireless networks considering many aspects of wireless communication and networking. For each aspect, we illustrate how machine learning can be applied using well-selected examples; and we pinpoint the advantages and disadvantages of using such machine learning/deep learning to solve each particular problem. We also provide some future research directions to overcome the challenges facing AI driven wireless communication and networking.

The rest of this paper is outlined as follows. Section II presents several applications of AI to solve issues in 5G wireless communication and networking. For reasons of space, the theory of machine learning and deep learning is not considered in this paper. Section III gives an overview on the challenges facing the integration of AI in 5G wireless networks as well as some future research direction to fully exploit AI in this context. Conclusions and future research directions are drawn in section IV.

## II. MACHINE LEARNING AND DEEP LEARNING

AI technology includes machine learning and deep learning. For the rest of this paper, we focus on deep learning because of the success deep learning have achieved. The theory of deep learning is comprehensively and well established. Nevertheless, for completeness reasons, we briefly give an overview of machine learning and deep learning. Machine learning techniques mainly can be classified into three main categories supervised learning, unsupervised learning, and reinforcement learning. In the first category, there is a mapping between the input and output. The machine learning models are given the labels of the dataset at the output, and it has to optimize the weights of the cost function so it can best learn the representations of the input data and the rules that map these inputs and their outputs. Examples of techniques under this category include logistic regression, support vector machine, decision tree, and random forest. In contrast, in the second category, the output's labels are not specified to the machine learning models, which itself has to underline any hidden patterns in the input and cluster the elements of the input dataset. Thus, it can be said that the primary function of unsupervised learning is underlying patterns instead of mapping the input and its labels. Examples of techniques under this category are clustering techniques such as K-means and self-organizing maps. In both supervised and unsupervised learning, there is no reward function, which is present in reinforcement learning that defines reward mechanisms to give feedback to the model. The last type is reinforcement learning, which is built upon establishing a reward mechanism. Similar to supervised learning, in reinforcement learning, there is a mapping between the input and the output. Over the past decade, a class of techniques called deep learning, which can be either supervised, unsupervised or reinforced, has been used in many technologies. Deep learning can be defined as a model which involves many hidden layers between the input layer and the output layer. Deep learning reveals unknown correlations in large data sets by using the feed-forward and back-propagation algorithms. One popular class of deep learning is convolutional neural network. A neural network is a network of neurons that are interconnected, and every neuron consists of a weighted sum of the inputs and one activation function, such as sigmoid function, rectified linear unit (RELU), threshold, and softmax. The main foundations on which neural networks are built are feed-forward propagation and backward-propagation algorithms. The first calculates the output as a function of the inputs. The latter computes the weights to minimize the error between the output predicted and the real one.
I
## III. WIRELESS AI

In this section, we selected several use cases in 5G wireless communication and networking empowered by machine learning and deep learning. For each example, we show the concepts, the advantages, and disadvantages of AI-enabled methodologies.

### A. Massive MIMO and Beamforming

Massive MIMO is one feature of 5G. Through the use of a vast number of antennas, 5G can focus the transmission and reception of signal power into ever-smaller regions of space. However, several issues are related to this technology. Machine learning/deep learning has been applied in Massive MIMO to overcome these issues. For instance, an accurate estimate of the channel with simple estimation methods and a reasonable number of pilots is challenging in massive MIMO: the low complexity least-squares (LS) estimator does not achieve satisfactory performance, while minimum mean square error (MMSE) channel estimation is very complex.

Machine learning/ deep learning can be used to bypass this issue. For instance, The authors of [2]–[9] proposed deep learning for channel estimation. Deep learning can be used also for symbol detection in MIMO systems as proposed in [10], [11] as it can be used for mapping channels in space and frequency as shown in [12]. Deep learning can be used for power allocation in Massive MIMO, as studied in [13].

Machine Learning and Deep learning have been also investigated in optimizing the weights of antenna elements in massive MIMO. Deep learning and machine learning can predict the user distribution and accordingly optimizing the weights of antenna elements, can improve the coverage in a multi-cell scenario [14]. Deep learning in massive MIMO presents several advantages. For instance, deep learning can carry out a more accurate channel estimation of the state of the channel compared to traditional techniques. Another advantage of using deep learning in this context is the reduction of the number of pilots required to achieve satisfactory performance. As a conclusion, massive MIMO can benefit from the power of deep learning if the complexity has been handled correctly.

### B. Automatic Modulation Classification

Automatic modulation classification (AMC) is a core technique in non-cooperative communication systems. Modulation recognition is one task that can help in classifying the modulation type of a received signal, which is a necessary step towards understanding and sensing the wireless environment. High-quality sensing and adaptation improve spectral efficiency and interference mitigation.

Deep learning-based AMC systems consist of three main parts: The first part is signal processing to enhance the quality of the received samples, a frequency offset correction, gain control, amplifiers, and filtering. The second part involves the extraction of features such as the amplitude, phase, and frequency of the received signal. The last part is a signal classifier: classification of the modulation types. Deep learning can achieve high accuracy of modulation classification. For instance, the authors of [15] proposed an ANN (two hidden layers with 50 and 25 neurons) that uses a system based AMC, which consists of Nesterov accelerated adaptive moment estimation algorithm to improve the training runtime. The authors transmitted several modulation techniques such as BPSK, QPSK,8PSK,16QAM, CPFSK, GFSK, and GMSK with different levels of power of SNR, ranging from 5dB to 45 dB. The collected signals have been preprocessed, and features such as amplitude, phase, frequency, signal statistics such as moments and cumulants have been extracted. The proposed model yields near-real-time modulation classification with an accuracy of 98%, but it is ineffective at low SNR.

The authors of [16] showed that long short term memory (LSTM) could achieve a classification accuracy close 90% at varying signal-to-noise ratio conditions ($0dB$ to $20dB$). The authors used the RadioML2016.10a dataset. The authors showed that LSTM with two layers outperforms support vector machine, random forest, naive Bayes, and K-nearest neighbors. Nevertheless, the performance of these models is less than 20% at SNR below $-10dB$.

The authors of [17] used AlexNet, which is a large CNN based Model that has eight convolution layers and three fully connected layers, to classify 11 modulation types which can achieve an average accuracy of 87%. AlextNet is developed to classify images, so the authors proposed the conversion of the complex samples of the modulated signal to a constellation diagram. They generated 10,000 images and 1000 constellation diagram images per modulation type. Each image consists of 1000 samples of the modulated signal, and the SNR ranges from $-4dB$ to $14dB$. AlexNet outperforms support vector machine and cumulant based AMC, and it does not require any feature selection step.

Deep learning presents a robust methodology for automatic modulation classification. This methodology has several advantages, such as the short processing time and the steady performance under low signal-to-noise ratio.

*C. Channel Coding*

A noticeable feature of the air interface of the 5G is the use of new channel coding techniques: Data channels use low-density parity-check (LDPC) codes, and control channels use polar codes [18]. However, the use of these techniques have some limitations. For instance, polar codes can achieve excellent performance, but it takes several iterations to achieve this performance, and there is no way to predict how fast polar codes can reach this desired performance. In addition, LDPC codes suffer from high complexity of decoding when either it is used with large block or the channel is under colored noise.

Deep learning is well-known for its high parallelism structure, which can implement one-shot coding/decoding. Thus, many researchers predict that deep learning-based channel coding is a propitious method to enable 5G NR. For instance, the authors of [19] proposed reinforcement learning for effective decoding strategies for binary linear codes such as ReedMuller and BCH codes, and as a case study, they considered bit-flipping decoding. The authors mapped learned bit-flipping decoding to a Markov decision process and reformulated the decoding problem using both standards and fitted Q-learning with a neural network. The neural network architecture consists of two hidden layers with 500 and 1500 neurons with ReLu activation functions. For the training hyperparameters, the authors considered ten iterations and 0.99 as a discount factor. The SNR is ranging from $-2dB$ to $8dB$. The authors considered two types of channels, binary symmetric channel, and Additive White Gaussian Noise (AWGN) channel.

The authors of [20] proposed three types of deep neural networks for channel decoding for 5G, multi-layer perceptron, convolutional neural network, and recurrent neural network. The authors used polar codes with rate 1/2 and three codeword lengths 8, 16, and 32. The signal to noise ratio is from $-2\ dB$ to $20\ dB$. The authors showed that the recurrent neural network has the best decoding performance but at the cost of high computation time.

The authors of [21] studied a low latency, robust, and scalable convolutional neural network-based decoder of convolutional and LPDC codes. The convolution decoder is trained to decode in a single-shot using Mixed-SNR independent sampling. The CNN decoder is tested with different block lengths of 100, 200, and 1000 under the AWGN channel and with total samples of $10^9$ samples, and SNR is ranging from $-4dB$ to $4dB$. The proposed model is compared with Viterbi, BiGRY, and bit flipping based decoders using bit error rate and block error rate. The authors showed that CNN outperforms the previously mentioned decoders regarding BER and BLER.
Also, CNN decoder is eight times faster than RNN decoders.

Another example of deep learning-based channel decoder is proposed in [22]. The proposed deep learning models consists of an iterative belief propagation concatenated with a convolutional neural network (BP-CNN) LDPC decoding under correlated noise, CNN for denoising the received signal and BP for decoding. The authors considered the AWGN channel and BPSK modulation. The authors showed that BP-
CNN reduces the decoding bit error rate with low complexity.

Further studies are required to investigate the performance of deep learning under communication channels which exhibit correlations in fading. Deep learning-based channel coding can achieve a good range of performance–complexity trade-offs, if the training is performed correctly as the choice of code-word length, causes over-fitting and under-fitting.

*D. Intelligent Radio Resource and Network Management*

Radio resources are scarce, and there is an increasing demand of wireless traffic. Intelligent wireless network management is the way forward to meet these increasing demands. Machine learning/deep learning can be a promising feature for resource allocation in 5G wireless communication networks. Deep learning can be a good alternative for interference management, spectrum management, multi-path usage, link adaptation, multi-channel access, and traffic congestion. For instance, the authors of [23] proposed an AI scheduler to infer the free slots in a multiple frequencies time division multiple access to avoid congestion and high packet loss. Four last frames state are fed to a neural network, which consists of two fully connected hidden layers. The proposed AI scheduler was tested in a wireless sensor network of 5 nodes and can reduce the collisions with other networks with 50%.

The authors of [24] proposed the addition of the artificial intelligence module instead of replacing conventional scheduling module in LTE systems. This AI module can provide conventional scheduling algorithms with the flexibility and speed up the convergence time. As scheduling for cooperative localization is a critical process to elevate the coverage and the localization precision, the authors of [25] presented a deep reinforcement learning for decentralized cooperative localization scheduling in vehicular networks.

The authors of [26] proposed a deep reinforcement learning (DRL) based on LSTM to enables small base stations to perform dynamic spectrum access to an unlicensed spectrum. The model enables the dynamic selection of wireless channel, carrier aggregation, and fractional spectrum access. The coexistence of WLAN and other LTE-LAA operators transmitting on the same channel is formulated as a game between the two and each of which aims to maximize its rate while achieving long-term equal-weighted fairness. This game is solved using DRL-LSTM. The proposed framework showed significant improvement.

The authors of [27] proposed an AI framework for smart wireless network management based on CNN and RNN to extract both the sequential and spatial features from the raw signals. These features serve as a state of deep reinforcement learning which defines the optimal network policy. The proposed framework was tested using real-experiment an experiment using a real-time heterogeneous wireless network test-bed. The proposed AI framework enhances the average throughput by approximately 36%. However, the proposed framework is costly in terms of training time and memory usage.

The authors of [28] proposed a deep-reinforcement learning approach for SDN routing optimization. To evaluate the performance of the proposed DRL based routing model, the scalefree network topology of 14 nodes, and 21 full-duplex links, with uniform link capacities and average node degree of 3, and traffic intensity levels from 12.5% to 125% of the total network capacity. The trained DRL routing model can achieve similar configurations that of methods such as analytical optimization or local-search heuristic methods with minimal delays. Some other work on routing can be found in [29], [30].

Another aspect of network management is interference management. Interference management often relays on algorithms such as WMMSE. This algorithm is costly as it uses matrix inversion, to solve the problem of numerical optimization in signal processing, the authors of [31] proposed to approximate the WMMSE used for interference management, which is has a central role in enabling Massive MIMO systems. The authors showed that SP optimization algorithms could be approximated by a finite-size neural network.

*E. Energy Efficiency Maximization*

Information and Communication Technology (ICT) is responsible for 2% to 10% of the world energy consumption in 2007, and it is expected to continue to grow [32]. Also, more than 80% of ICT is from radio access network (RAN), which is deployed to meet the peak traffic load and stays on it even that the load is light. Motivated by saving energy for green communication, 5G specifications require that energy use should decrease to 10% percent of the traditional 4G/LTE networks. This objective can be achieved by reducing the power consumption of the base stations and mobile devices.

Many researchers investigated the use of deep learning theory to minimize the energy consumption in 5G wireless networks [33]–[36]. For instance, the authors of [33] proposed a deep reinforcement learning-based Small cell base stations (SBSs) activation strategy to lower the energy consumption without comprising the quality of service. In particular, the SBS on/off switching problem is formulated into a Markov decision process and solved by actor-critic (AC) DRL. Energy consumption of the network alongside the quality of service degradation and mode switching costs are the cost metrics in this study. The networks have two hidden layers with a number of neurons of 200 and 100, respectively. The model is trained, and the daily cost is the average of 20 days cost of 20 instances. The authors of [34] data-driven base station sleeping operations through deep reinforcement learning.

Machine learning/ deep learning thus can help in building intelligent wireless networks that proactively predict the traffic and mobility of users and delivery services only when requested — subsequently reducing the power consumption in radio access networks. The authors of [36] developed a deep learning power control framework for energy efficiency maximization in wireless interference networks. Throughout the above-mentioned examples, deep learning can reduce energy consumption in 5G radio access networks.

*F. 5G Slicing and Caching*

Two prominent features of 5G are the network slicing and caching. The first allows operators to deliver different service types over the one network infrastructure. The latter predicts the content that users may request for efficient usage of the storage of the base station. Thus, the 5G requires accurate predictions of the needed resources in a slice and the future content of the users.

Several research works have investigated 5G resource provisioning and caching using the theory of machine learning/ deep learning. For example, the authors of [37] proposed X-LSTM to predict future usage to manage 5G slicing. A metric called REVA is developed, and to forecast REVA the next 30 seconds with prediction intervals of 5 seconds. The authors developed X-LSTM, which is built upon LSTM and ARIMA, which a popular statistical method. This methodology enables an improvement of X-LSTM over ARIMA and LSTM for X-LSTM outperformed the other time series models by 10%, 22%, and 31%, respectively. Also, X-LSTM results in more than 10% cost reduction per slice. The authors of [38] proposed a novel caching framework for offloading the back-haul and front-haul loads in a CRAN system. The proposed algorithm enables the prediction of the content request distribution of each user with limited information on the network state and user context.

*G. 5G Cybersecurity*

Deep learning has also been investigated in cybersecurity of 5G wireless communications. For instance, the authors of [39] proposed an unmanned aerial vehicle (UAVs) aided 5G wireless communications with deep reinforcement learning against jamming attacks. The relay UAVs are used to establish the communication of legitimate nodes. To determine the optimal policy of the relay UAV, the authors addressed proposed a deep reinforcement learning. The methodology can restore the communication between the base station and the legitimate users, but several issues need to be addressed to enable these anti-jamming methods. The authors of [40] investigated the robustness of deep learning in wireless communication systems against physical adversarial attacks. The authors of [41] proposed a machine learning model for power control for mmWave Massive MIMO against jamming attacks. The authors of [42] proposed a 5G cyber-defense architecture to identify cyber-threats in 5G wireless networks. This defense architecture uses deep learning to inspect the network traffic by extracting features from the traffic flow. In particular, the authors used LSTM, which is trained on the CTU dataset that is a public dataset that contains real-traffic.

IV. DISCUSSION AND FUTURE RESEARCH DIRECTIONS

In this section, we revisit the advantages of deep learning and machine learning in building intelligent 5G wireless communication and networking. We discuss also the challenges facing the integration of AI in wireless communication as well as some future research directions to speed up this integration.

*A. Advantages*

The use of deep learning to build intelligent 5G systems has many advantages. For instance, in signal processing, deep learning is capable of performing automatic feature extraction, which is a hard task in wireless network engineering that often requires human expertise. Deep learning can perform this task with high accuracy. Another advantage is that deep learning models, in some cases, can achieve high-performance accuracy and outperform traditional techniques. 5G wireless networks are expected to generate a huge amount of data at high data rates. Deep learning can enable 5G systems to take advantage of this to engineer optimized wireless networks.

*B. Challenges*

The integration of AI in 5G wireless communication systems faces many challenges. Some of these challenges can be listed as follows:

*1) The reliability and speed trade-off:* The reliability of these techniques is far less than traditional techniques in wireless communications in solving some problems. For instance, deep learning can compete with LS and MMSE in wireless channel estimation in massive MIMO, but slow feedback characterizes these techniques. Deep learning inference may elongate the system response time. This is because not most wireless devices have access to cloud computing, and even if it is the case, communication with cloud servers is going to introduce extra delays.

*2) The complexity:* Deep learning algorithms, in due course, need to be implemented in wireless devices. However, many wireless devices have limited memory and computing capabilities, which is not suitable for complex algorithms. The collection of large samples and training deep learning models takes considerable time, which is a significant impediment to deploy them on some wireless devices having limited power and storage. Also, some applications require real-time processing, and on-fly sampling and training often cannot be performed easily. In some cases, the higher the number of samples and the more significant the training time are, the higher the accuracy of recognition of the signal and network features is. Acquiring more samples and training the models for longer times incur slow feedback. Therefore, the deep learning models should be designed to achieve the best accuracy with fewer samples and within a short time.

*3) Data Collection and Cleansing:* It is necessary to collect data and build large comprehensive datasets to train AI models, and this task is not often easy to acquire because mobile service providers, for instance, cannot release these datasets, which contains confidential information about the users and can risk the violation of the privacy of their consumers. Also, even with transfer learning, which refers to use models trained on the previous dataset, it is necessary to adapt these models for specific networks and scenarios which require re-training of the models. All these reasons restrict the development of wireless AI

*4) Privacy:* Preserving the privacy of the users is the primary concern of mobile and service providers. One of the main challenges in wireless AI is how one can enable the training on a dataset belonging to users without sharing the input data and putting the personal information of users at risk. It is necessary to have a security approach to boost the integration of deep learning in wireless communications.

*5) Security:* The security of deep learning models itself in another challenge, as neural networks are prune to adversarial attacks. Attackers can affect the training process by injecting fake training datasets; such injection can lower the accuracy of the models and yield wrong design, which may affect the network performance. Research in the security of deep learning or machine learning, in general, remains shallow.

*C. Future Research Directions*

In order to ease the integration of deep learning, research efforts are needed in several directions. For instance, the acceleration of deep neural network alongside advanced parallel computing, faster algorithm, and cloud computing, distributed deep learning systems present an opportunity for 5G to build the intelligence in its systems to deliver high throughput and ultra-low latency. There have been some recent efforts in deep neural network acceleration [43]. The acceleration of deep neural network, can be at three levels: architecture level, computation level, and implementation level. At the architecture level methods can be used, including layer decomposition [44], pruning [45], projection [46], and knowledge distillation [47]. At the implementation level, several characteristics can be explored such as advanced GPU [48] and FPGA designs [49]. Using deep learning acceleration methods can achieve lower the complexity of deep learning with small loss in the accuracy of these models. Combining these methods can reduce the number of parameters by more than 50%. Further exploration of the acceleration of these networks can have a huge impact on the adoption of this deep learning to build intelligence in 5G systems.

Another way to speed up the integration of deep learning theory is 5G wireless communication systems is data collection and cleansing as there are not many datasets available so researchers that can used to build and test their models. Efforts in this directions are highly needed to build systems that can generate dataset.

## V. CONCLUSION

In this paper, we presented AI for 5G wireless communication systems. We studied several case studies including modulation classification, channel coding, massive MIMO, caching, energy efficiency, and cybersecurity. As a conclusion of this in-depth study, AI enabled 5G wireless communication and networking is a promising solution that can provide wireless networks with the intelligence, efficiency, and flexibility required to manage the scare radio resource well and deliver high quality of service to the users. However, some efforts are still needed to reduce the complexity of deep learning so it can be implemented in time-sensitive networks and low power devices and test the models in more realistic scenarios.